\documentclass[prd,twocolumn,showpacs,superscriptaddress,nofootinbib,preprintnumbers,aps]{revtex4}
\usepackage{color}
\usepackage{latexsym, amssymb, amsmath,color,graphicx}
\usepackage[unicode=true, 
 bookmarks=true,bookmarksnumbered=false,bookmarksopen=false,
 breaklinks=false,pdfborder={0 0 1},backref=false,colorlinks=true]
 {hyperref}

 \makeatletter

\@ifundefined{textcolor}{}
{%
 \definecolor{BLACK}{gray}{0}
 \definecolor{WHITE}{gray}{1}
 \definecolor{RED}{rgb}{1,0,0}
 \definecolor{GREEN}{rgb}{0,1,0}
 \definecolor{BLUE}{rgb}{0,0,1}
 \definecolor{CYAN}{cmyk}{1,0,0,0}
 \definecolor{MAGENTA}{cmyk}{0,1,0,0}
 \definecolor{YELLOW}{cmyk}{0,0,1,0}
 }

\usepackage{hyperref}
\hypersetup{
    colorlinks,%
    citecolor=blue,%
    filecolor=blue,%
    linkcolor=blue,%
    urlcolor=blue
}

\newcommand{\p}{\partial}

\begin{document}

\preprint{CERN-PH-TH/2011-47}

\title{Non-Stationary Dark Energy Around a Black Hole}

\author{Ratindranath Akhoury}
\email{akhoury@umich.edu}
\affiliation{Michigan Center for Theoretical Physics, Randall Laboratory of Physics, University of Michigan, Ann Arbor, MI 48109-1120, USA}

\author{David Garfinkle}
\email{garfinkl@oakland.edu}
\affiliation{Department of Physics, Oakland University, Rochester, MI 48309, USA}
\affiliation{Michigan Center for Theoretical Physics, Randall Laboratory of Physics, University of Michigan, Ann Arbor, MI 48109-1120, USA}

\author{Ryo Saotome}
\email{rsaotome@umich.edu}
\affiliation{Michigan Center for Theoretical Physics, Randall Laboratory of Physics, University of Michigan, Ann Arbor, MI 48109-1120, USA}

\author{Alexander Vikman}
\email{alexander.vikman@cern.ch}
\affiliation{Theory Division, CERN, Geneva 23, CH-1211, Switzerland}

\date{\today}

%%%%%%%%%%%%%%%%%%%%%%%%%%%%%%%%%%%%%%%%%%%%%%%%%%%
\begin{abstract}

Numerical simulations of the accretion of test scalar fields with non-standard kinetic terms (of the k-essence type) onto a Schwarzschild black hole are performed. We find a full dynamical solution for the spherical accretion of a Dirac-Born-Infeld type scalar field. The simulations show that the accretion eventually settles down to a well known stationary solution. This particular analytical steady state solution maintains two separate horizons. The standard horizon is for the usual particles propagating with the limiting speed of light, while the other sonic horizon is for the k-essence perturbations propagating with the speed of sound around this accreting background. For the case where the k-essence perturbations propagate superluminally, we show that one can send signals from within a black hole during the approach to the stationary solution. We also find that a ghost condensate model settles down to a stationary solution during the accretion process.

\end{abstract}

%%%%%%%%%%%%%%%%%%%%%%%%%%%%%%%%%%%%%%%%%%%%%%%%%%%%%%%%%%%%%%%%%%%%%%%%%%%%%%

\maketitle

%%%%%%%%%%%%%%%%%%%%%%%%%%%%%%%%%%%%
%%%%%%%%%%%%%%%%%%%%%%%%%%%%%%%%%%%%
\section{Introduction}

Recent observations in cosmology suggest that the universe is expanding at an accelerated rate. In order to explain this acceleration, the energy density of the universe must have a negative pressure dark energy component. This dark energy component is most commonly modeled as a cosmological constant. However, this explanation fails to address why the universe has entered an accelerating era shortly after the onset of matter domination. One dark energy model that addresses this issue is k-essence \cite{steinhardt,steinhardt2}, which postulates that there is a dynamical scalar field with non-canonical kinetic terms that acts as a negative pressure component only after the onset of matter domination. In this context it is worth noting that in order to solve this coincidence problem there must exist a short phase of superluminal sound speed of the k-essence fluctuations \cite{durrer, Kang:2007vs}. However, this faster than light propagation does not contradict causality, see e.g. \cite{Babichev:2007dw} and references therein. In order to determine if nature indeed utilizes such scalar fields we must look for unique signatures of k-essence models. Some possible areas worth investigating are the gravitational collapse of k-essence \cite{ags} and the evolution of black holes surrounded by various non-canonical scalar fields \cite{bmv,bmv2,babichev,frolov,mukho,Babichev:2008dy,Babichev:2010kj}.

Numerical simulations \cite{ags} have shown that the gravitational collapse of k-essence fields proceeds  through an intermediate stage of a black hole with two separate horizons, one being the traditional event horizon that corresponds to the speed of light, and the other being a sonic horizon which corresponds to the propagation speed of k-essence field perturbations. As these previous simulations required asymptotic flatness, only k-essence shells with finite width could be simulated.

Because we know that these k-essence fields can form black holes, an interesting question to answer is whether or not these fields eventually settle down and have simple behavior during accretion onto a Schwarzschild black hole. The simplest set-up for analytical studies of accretion is the steady state or Bondi accretion \cite{bondi}. It was shown in \cite{agv} that for general k-essence fields exact stationary configurations are possible only for shift symmetric theories \footnote{For time-like gradients these shift-symmetric theories are equivalent to barotropic perfect  fluids , see e.g. \cite{Arroja:2010wy, Unnikrishnan:2010ag,Moncrief} }. Since the Lagrangian has to be invariant under constant translations in field space, it is a function only of the derivatives of the scalar field. For such Lagrangians exact stationary solutions have been obtained \cite{babichev,bmv,bmv2,frolov,mukho} and steady state accretion analyzed. Indeed the existence of two separate horizons, the usual light horizon and a sound horizon which traps k-essence fluctuations was to our best knowledge first discussed  in \cite{Moncrief} for general irrotational hydrodynamics. While this is very interesting, we would really like to know the outcome of a full dynamical evolution as the scalar field accretes on to the already existing black hole. In this case there will be a constant influx of scalar fields onto the black hole and it is far from clear that at late times a stationary or steady state configuration is reached. Because there is a 
constant influx of energy, a steady state is only one possibility.  Alternatively, the evolution of the scalar field
could result in a singularity, or the end state could be an oscillatory solution rather than a stationary one (as happens for 
variable stars). In particular it was shown in \cite{bmv} that for a superluminal DBI-type model and sound speeds higher than $2/\sqrt{3}$ at spatial infinity a steady state solution suffers from catastrophic gradient instabilities indicating that it is a repeller. 

Stability of the steady state accretion is well studied, see e.g. \cite{Moncrief} in the case of the time-like gradients when the shift-symmetric theory is equivalent to hydrodynamics. Since we are studying the full dynamical evolution of the scalar field, there can be regions where the gradient of the scalar field is space-like even though we start with a time-like gradient in our initial conditions. This has no effect on the final state of the scalar field, which we find is always stationary.

In order to resolve some of these issues, in this paper we will simulate the accretion of a test k-essence field. In the case that the end behavior of this k-essence scalar field during the accretion process is stationary (if the field settles down at all), we will be able to compare our end result to the stationary analytic solutions found in the literature \cite{bmv,frolov}. This will require us to use cosmological boundary conditions, where at infinite radius we set the time derivative of the k-essence field to a constant. The two separate horizons can exist for a long enough time to confirm if signals can escape from the black hole when the sound horizon is inside the usual black hole horizon.

The paper is organized as follows: In the next section we discuss the method used for the simulations and the results are presented in section 3. Certain technical details of the construction of the exact stationary solutions for the models considered are relegated to an appendix.

\section{Methods}

The action for a k-essence scalar field in a background curved spacetime is 
\begin{equation}
I = \int {\sqrt g} \, {d^4} x \, {\cal L}(X) \ ,
\end{equation}
where $ X = - {\frac 1 2} g^{\mu\nu}{\nabla _\mu} \phi {\nabla _\nu} \phi$ is the standard kinetic term for the scalar field $\phi$ and $g$ is the absolute value of the determinant of the metric $g_{\mu\nu}$. 
The use of a fixed background means that we are neglecting the gravitational effects of the scalar field. 
Varying the action with respect to $\phi$ yields the equation of motion
\begin{equation}
{{\tilde g}^{\mu \nu}} {\nabla _\mu}{\nabla _\nu} \phi = 0 \ ,
\label{kefe}
\end{equation}
with ${\tilde g}^{\mu \nu}$ being an effective inverse (contravariant) metric associated with the k-essence acoustic geometry and is given by
\begin{equation}
{{\tilde g}^{\mu \nu}} = {{\cal L}_X}  {g^{\mu \nu}} - {{\cal L}_{XX}} {\nabla ^\mu}\phi {\nabla ^\nu}\phi
\label{geff} \ .
\end{equation}
Here, we introduce the shorthand ${\cal L}_X$ to mean $d{\cal L}/dX$ and
${\cal L}_{XX}$ to mean ${d^2}{\cal L}/d{X^2}$.

We simulate the evolution of k-essence on a background Schwarzschild spacetime.  In the usual coordinates, the 
Schwarzschild metric is given by
\begin{equation}
d {s^2} = - f(r) d {t^2} + {f^{-1}}(r) d {r^2} + {r^2} (d {\theta ^2} + {\sin ^2} \theta d {\varphi ^2})\ ,
\end{equation}
where $f(r)=1-(2M/r)$.  
However, this coordinate system is not suitable for our simulation, since it becomes singular on the black hole horizon.  The usual
solution to this difficulty is to introduce 
Eddington-Finkelstein coordinates ($v,r,\theta,\varphi$) where $v$ is defined by
\begin{equation}
v = t + \int {f^{-1}} d r\ .
\label{coords1}
\end{equation}
This coordinate system covers a region that includes the black hole interior; but it
also is not suitable for our purposes since $v$ is a null coordinate, and as with most numerical methods we require a
time-like coordinate.  Instead, we use the method of Marsa and Choptuik \cite{matt}. Introduce the coordinate $T$ by 
\begin{equation}
T = v - r \ .
\label{coords2}
\end{equation}
Then the Schwarzschild metric takes the form
\begin{eqnarray}
d {s^2} = &-& \left ( 1 - {\frac {2 M} r} \right ) d {T^2} +  \left ( 1 + {\frac {2 M} r} \right ) d {r^2} 
\nonumber
\\
&+& {\frac {4 M} r} dT dr 
+ {r^2} (d {\theta ^2} + {\sin ^2} \theta d {\varphi ^2}) \ .
\label{schmatt}
\end{eqnarray}
In order to write the k-essence equation of motion in first order form, we introduce the quantites $P$ and $S$ defined by
\begin{eqnarray}
P = {\partial _T} \phi \ ,
\label{Pdef}
\\
S = {\partial _r} \phi \ .
\label{Sdef}
\end{eqnarray}
Eqn. (\ref{Pdef}) provides an equation of motion for $\phi$ while from eqn. (\ref{Sdef}) it follows that
\begin{equation}
{\partial _T} S = {\partial _r} P \ ,
\label{dtS}
\end{equation}
which is an equation of motion for $S$.  Finally, some straightforward but tedious algebra using eqns. (\ref{kefe}) 
and (\ref{schmatt}) yields the equation of motion for $P$ which takes the form
\begin{equation}
{\partial _T} P = {\frac {{{\cal L}_X} A + {{\cal L}_{XX}} B} C} \ ,
\label{dtP}
\end{equation}
where the quantities $A, \, B$ and $C$ are given by  
\begin{eqnarray}
A &=& {\frac {4M} r} {\partial _r} P + \left ( 1 - {\frac {2M} r} \right ) {\partial _r} S + {\frac {2S + w} r} \ ,
\\
B &=&  (S+w) \biggl [ 2(P+w) {\partial _r} P 
\nonumber
\\
&-& (S+w){\partial _r} S - {\frac {w^2} {4M}}\biggr ] \ ,
\\
C &=& \left ( 1 + {\frac {2M} r}\right ) {{\cal L}_X} 
+ {{(P+w)}^2} {{\cal L}_{XX}} \ ,
\label{Cdef}
\end{eqnarray}
and the quantity $w$ is given by
\begin{equation}
w = {\frac {2M} r} (P-S) \ .
\end{equation}

We numerically evolve eqns. (\ref{Pdef}), (\ref{dtS}), and (\ref{dtP}) as follows: time derivatives are treated
using the iterated Crank-Nicholson method.  
Spatial derivatives, except those on the boundary, are evaluated using standard 
centered differences; that is, for any quantity $F$ we approximate ${\partial _r}F$ at grid point $i$ by
\begin{equation}
{\frac {{F_{i+1}} - {F_{i-1}}} {2 d r}} \ ,
\end{equation}
where $dr$ is the spacing between adjacent grid points.  At the boundary, spatial derivatives are evaluated using
one-sided differences: that is at grid point $1$ we approximate ${\partial _r}F$ by
\begin{equation}
{\frac {{F_2} - {F_1}} {dr}} \ ,
\end{equation}  
while at the final gridpoint (point $N$) we approximate ${\partial _r}F$ by
\begin{equation}
{\frac {{F_N} - {F_{N-1}}} {dr}} \ .
\end{equation}

Physically appropriate boundary conditions must also be imposed on the k-essence field at each step of the time evolution.  
We are treating an overall constant flux of scalar field along with waves of scalar field that should be outgoing at
the outer boundary.  Therefore we impose the condition 
\begin{equation}
P + S = {c_p} \ ,
\end{equation}
at the outer boundary, where $c_p$ is a constant representing the rate of flux of scalar field.  At the inner
boundary, we use an excision method appropriate for k-essence.  That is, we place the inner boundary sufficiently
far inside the black hole that not only light but also k-essence modes are trapped and therefore ingoing at
that point.  Since all modes are ingoing at the inner boundary, no boundary condition is needed (or even allowed)
there, and therefore we do not impose any boundary condition at the inner boundary.

\section{Results}

Unless stated otherwise, for all results and figures shown we chose the initial data at $T=0$ such that $\phi=0$, $S=0$, and $P=0.01$. The mass $M$ of the black hole was chosen to be unity, so the corresponding standard Schwarzschild horizon is located at $r=2$ for all shown simulations.

\subsection{DBI-Type Action}

\begin{figure}
\includegraphics{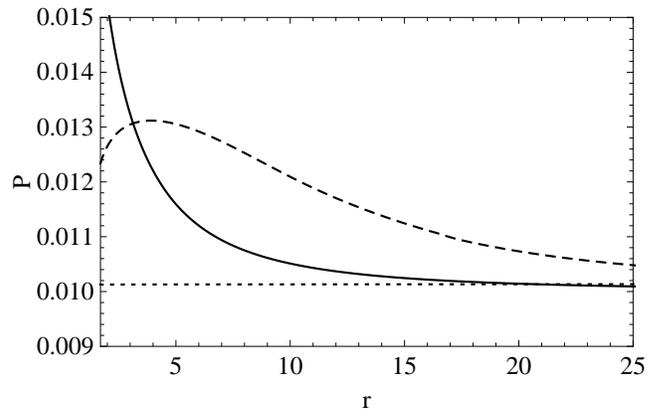}
\caption{Profiles of $P$ arrived at via simulation of the accretion of the field described by \eqref{Vik}. The solid line corresponds to $P$ at $T=2.97524$, the dashed line corresponds to $P$ at $T=14.8762$, and the dotted line corresponds to $P$ at $T=148.771$. For this particular plot, we used $c_{p}=0.01$, $\alpha=0.0016$, and $M=1$.}
\label{PosAPFig}
\end{figure}

\begin{figure}
\includegraphics{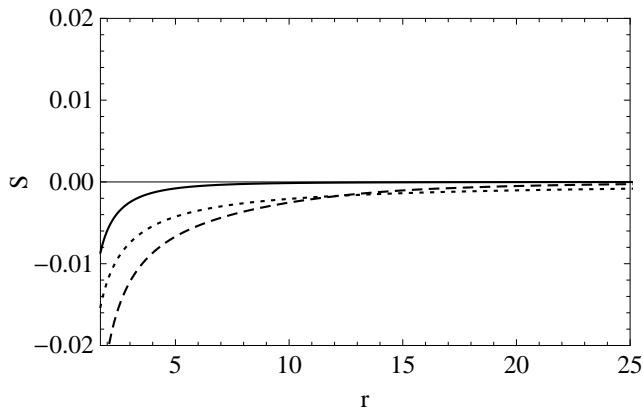}
\caption{Profiles of $S$ arrived at via simulation of the accretion of the field described by \eqref{Vik}. The solid line corresponds to $S$ at $T=2.97524$, the dashed line corresponds to $S$ at $T=14.8762$, and the dotted line corresponds to $S$ at $T=148.771$. For this particular plot, we used $c_{p}=0.01$, $\alpha=0.0016$, and $M=1$.}
\label{PosASFig1}
\end{figure}

We will first consider a k-essence scalar field with the DBI-like Lagrangian density:
\begin{align}
\mathcal{L}(X)=\alpha\left[\sqrt{1+\frac{2X}{\alpha}}-1\right] \ .
\label{Vik}
\end{align}
Note that the expression for the speed of sound (for time-like gradients) for this model is
\begin{align}
c_{\text{s}}^{2}=1+\frac{2X}{\alpha} \ .
\end{align}
Thus, for negative $\alpha$ the propagation is subluminal and for positive $\alpha$ the propagation is superluminal. Moreover, for time-like gradients this field theory reproduces one of the versions of a rather popular model among cosmologists, the so-called Chaplygin gas with the equation of state $p=-\alpha /  \rho$,  see  \cite{Kamenshchik:2001cp}. This form of the kinetic term with positive $\alpha$ was proposed for the first time in \cite{Mukhanov:2005bu}. The analytical solution for a steady state accretion  in this model was found in \cite{bmv, bmv2}. Upon simulating the accretion of this type of k-essence model into a Schwarzschild black hole, we found that for all positive $\alpha$ where the simulation ran successfully, the scalar field eventually settled down to a stationary solution. Simulated profiles for both $P$ and $S$ at various times are shown in figures \ref{PosAPFig} and \ref{PosASFig1}.

For sufficiently small values of positive $\alpha$ (for our initial conditions this is approximately $\alpha=0.001$), the simulation did not run successfully. This is because for these values of $\alpha$, at some point in the simulation the quantity $C$ given
in eqn. (\ref{Cdef}) goes to zero.  Since this quantity occurs as the denominator in eqn. (\ref{dtP}), the evolution equation
for $P$ becomes singular.  To understand what this means, note that $C$ is proportional to ${\tilde g}^{TT}=\mathcal{L}_{X}g^{TT}-\mathcal{L}_{XX}\nabla^{T}\phi\nabla^{T}\phi$. Thus, in physical terms, our time coordinate $T$ is no longer a valid global time coordinate in the emergent k-essence spacetime. This is because when  ${\tilde g}^{TT}=0$, we have:
\begin{align}
{\tilde g}^{\mu\nu}\nabla_{\mu}T\nabla_{\nu}T=0 \ ,
\end{align}
which indicates $\nabla_{\mu}T$ is not time-like in the emergent spacetime. This does not necessarily mean that there is no scalar function that can serve as a valid global time coordinate, which would indicate that the emergent spacetime is not stably 
causal \cite{wald}.

\begin{figure}
\includegraphics{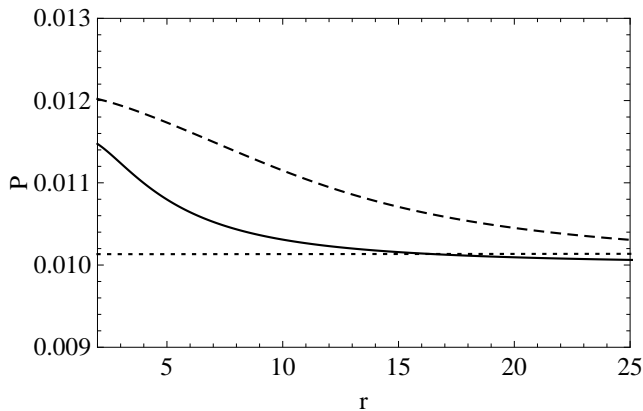}
\caption{Profiles of $P$ arrived at via simulation of the accretion of the field described by \eqref{Vik}. The solid line corresponds to $P$ at $T=2.97074$, the dashed line corresponds to $P$ at $T=14.8537$, and the dotted line corresponds to $P$ at $T=148.536$. For this particular plot, we used $c_{p}=0.01$, $\alpha=-0.0004$, and $M=1$.}
\label{MinAPFig}
\end{figure}

\begin{figure}
\includegraphics{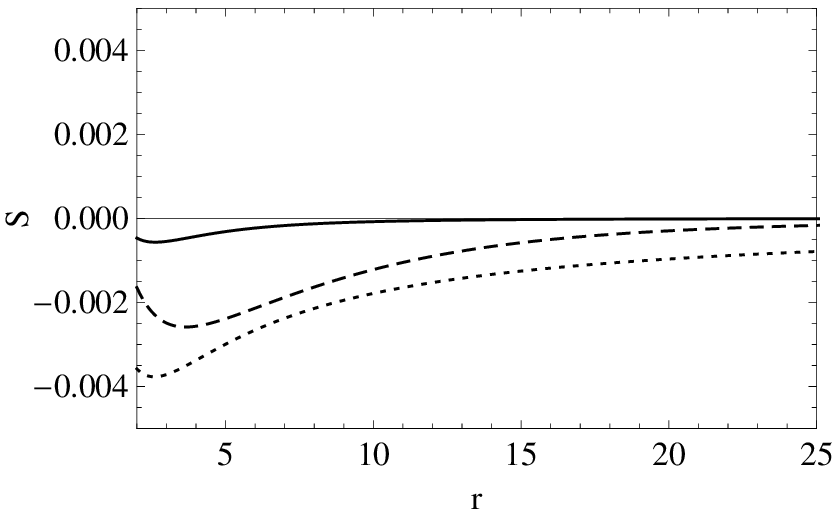}
\caption{Profiles of $S$ arrived at via simulation of the accretion of the field described by \eqref{Vik}. The solid line corresponds to $S$ at $T=2.97074$, the dashed line corresponds to $S$ at $T=14.8537$, and the dotted line corresponds to $S$ at $T=148.536$. For this particular plot, we used $c_{p}=0.01$, $\alpha=-0.0004$, and $M=1$.}
\label{MinASFig1}
\end{figure}

\begin{figure}
\includegraphics{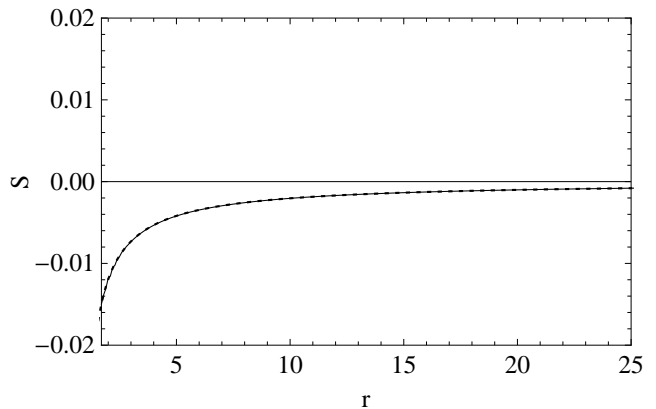}
\caption{The solid line is the radial derivative of the stationary solution for $\phi$ eventually arrived at via simulation of the accretion of the field described by \eqref{Vik}. The dotted line is the radial derivative of $\phi$ for the analytic stationary solution. For this particular plot, we used $c_{p}=0.01$, $\alpha=0.0016$, and $M=1$. The sound horizon has settled down at $r=1.88$. The radial derivative was chosen since the radial derivative of a stationary solution is time independent.}
\label{fig1}
\end{figure}

We also simulated this k-essence model for negative values of $\alpha$, since in this case we can ensure that ${\tilde g}^{TT}\ne0$. Simulated profiles for both $P$ and $S$ for $\alpha=-0.0004$ at various times are shown in figures \ref{MinAPFig} and \ref{MinASFig1}.  In this subluminal case we also find that for all successfully run simulations the scalar field settles down to a stationary solution. For values of negative $\alpha$ sufficently close to zero (for our initial conditions this is approximately $\alpha=-0.00015$), the simulation did not run successfully due to the speed of sound approaching zero. This is indicative of a shockwave, which our code was not meant to handle.

For both positive and negative $\alpha$, it was found that the scalar field would settle down to a stationary solution for a wide range of initial conditions as long as the same cosmological boundary condition ($P+S=c_{p}$) was maintained at infinite radius. An example of such inital data is:
\begin{align}
P=\frac{c_{p}}{2}(\mathrm{tanh}(r-r_{0})+1) \ ,
\end{align}
for various choices of $c_{p}$ and $r_{0}$, with $\phi=0$ and $S=0$.

\begin{figure}
\includegraphics{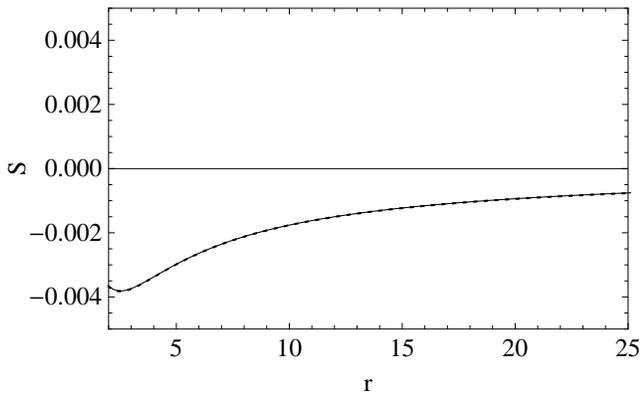}
\caption{The solid line is the radial derivative of the stationary solution for $\phi$ eventually arrived at via simulation of the accretion of the field described by \eqref{Vik}. The dotted line is the radial derivative of $\phi$ for the analytic stationary solution. For this particular plot, we used $c_{p}=0.01$, $\alpha=-0.0004$, and $M=1$. The sound horizon has settled down at $r=2.67$. The radial derivative was chosen since the radial derivative of a stationary solution is time independent.}
\label{fig2}
\end{figure}

In order to confirm that the fields did indeed settle down to a stationary solution, we compared our results to the stationary solution found analytically in \cite{bmv}. Expressed in our coordinates, the solution is (for a brief derivation see the appendix):
\begin{align}
\phi(T,r)=c_{p}\left(T+r+\int F(r)dr\right) \ ,
\end{align}
where
\begin{align}
F(r)\equiv\frac{1}{f(r)}\left(\sqrt{\frac{c_{\infty}^{2}+f(r)-1}{f(r)\frac{r^{4}c_{\infty}^{8}}{16M^{4}}+c_{\infty}^{2}-1}}-1\right) \ ,
\end{align}
 with $f(r)=1-(2M/r)$, and 
\begin{equation}
c_{\infty}^{2}\equiv c_{s}^{2}(r\rightarrow\infty)=1+\frac{c_{p}^{2}}{\alpha} \ .
\end{equation}
Comparisons between the analytic stationary solution and the ones arrived at via simulation for both the positive and negative $\alpha$ cases are shown in figures (\ref{fig1}) and (\ref{fig2}).

\begin{figure}
\includegraphics{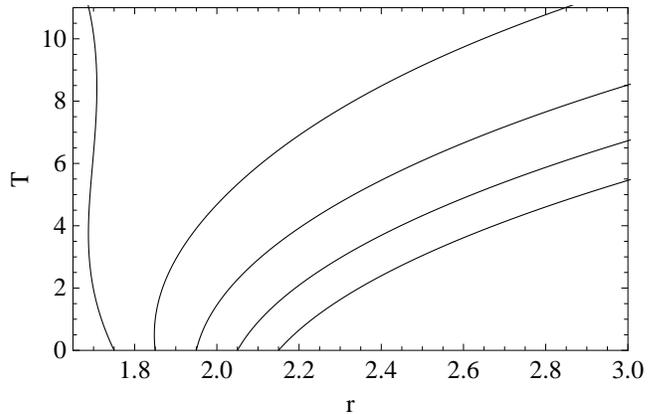}
\caption{Outgoing radial null geodesics corresponding to the emergent k-essence metric. For this simulation, we chose the values $c_{p}=0.01$, $\alpha=0.0016$, and $M=1$. We show five geodesics that start at $r=$ 1.75, 1.85, 1.95, 2.05 and 2.15. The geodesic starting at 1.75 fails to escape the black hole; the other four all escape, with two of the escaping geodesics starting within the Schwarzschild radius at $r=2$.}
\label{geodesic}
\end{figure}

We can define a sonic horizon as being the outermost radius at which null vectors with respect to the emergent k-essence metric become trapped. The null vectors $l^{\mu}$ are defined by:
\begin{align}
{\tilde g}^{-1}_{\mu\nu}l^{\mu}l^{\nu}=0 \ ,
\end{align}
where ${\tilde g}^{-1}_{\mu\nu}$ is the inverse of ${\tilde g}^{\mu\nu}$.
We can solve this quadratic equation for ${l^{r}}/{l^{T}}$, taking the positive root for outgoing null vectors. The sonic horizon is the outermost radius at which ${l^{r}}/{l^{T}}=0$. For stationary solutions corresponding to scalar fields with the Lagrangian density of \eqref{Vik}, the sonic horizon is located within the Schwarzschild radius for positive $\alpha$ and located outside the Schwarzschild radius for negative $\alpha$. To see that this implies that one can send signals from within a black hole for the superluminal positive 
$\alpha$ models, we should plot a set of radial null geodesics with respect to the emergent k-essence metric. Because for radial null geodesics ${dr}/{dT}={l^{r}}/{l^{T}}$, we can numerically integrate this quantity to find these geodesics, which are plotted in figure \ref{geodesic}. As can be seen in figure \ref{geodesic}, there exist null geodesics that start inside the Schwazschild radius at $T=0$ (when as stated earlier, $P=c_{p}$ and $\phi =0$ for all radii) and end up escaping as the field evolves towards the stationary solution. We want to emphasize that the null k-essence geodesics can escape a black hole not only in the stationary solution but also during the approach towards the stationary solution from more general initial conditions.

\subsection{Ghost Condensate Action}
\begin{figure}
\includegraphics{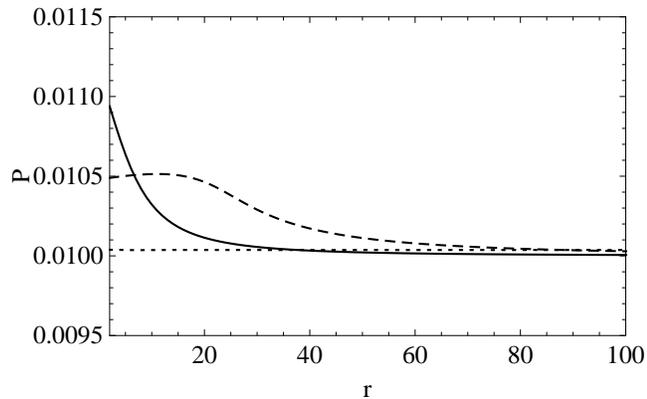}
\caption{Profiles of $P$ arrived at via simulation of the accretion of the field described by \eqref{ghost}. The solid line corresponds to $P$ at $T=11.9406$, the dashed line corresponds to $P$ at $T=59.7061$, and the dotted line corresponds to $P$ at $T=597.357$. For this particular plot, we used $c_{p}=0.01$, $a=0.0638943$ (which corresponds to a sound speed at infinite radius having the value $c_{\infty}=\frac{1}{2}$), and $M=1$.}
\label{UDMPFig}
\end{figure}

\begin{figure}
\includegraphics{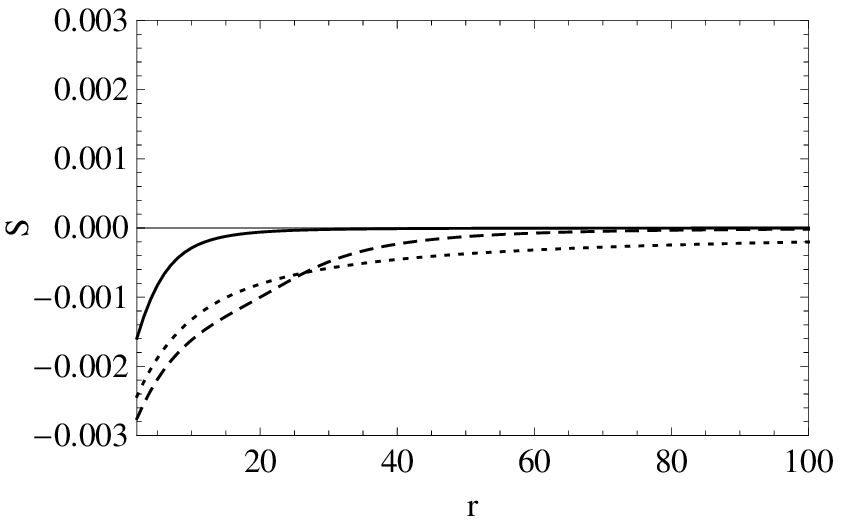}
\caption{Profiles of $S$ arrived at via simulation of the accretion of the field described by \eqref{ghost}. The solid line corresponds to $S$ at $T=11.9406$, the dashed line corresponds to $S$ at $T=59.7061$, and the dotted line corresponds to $S$ at $T=597.357$. For this particular plot, we used $c_{p}=0.01$, $a=0.0638943$ (which corresponds to a sound speed at infinite radius having the value $c_{\infty}=\frac{1}{2}$), and $M=1$.}
\label{UDMSFig1}
\end{figure}

Next we consider the ghost condensate models proposed in \cite{arkani}. These provide a modification of gravity in the infrared regime with possible ramifications for the dark matter problem and inflation. This model is a k-essence like model, where $X$ develops a non-zero vacuum expectation value. The ghost condensate field is not directly coupled to other fields, so the only scale in the ghost sector is $a$, the overall energy scale of the condensate.
The simplest such model is described by \cite{frolov,mukho}:
\begin{align}
\mathcal{L}(X)=\frac{1}{8a^{4}}(X-a^{4})^{2} \  .
\label{ghost}
\end{align}
The sound speed of the scalar field with this Lagrangian is given by:
\begin{align}
c_{s}^{2}=\frac{X-a^{4}}{3X-a^{4}} \ .
\end{align}
Thus, by making $|a|$ larger for a fixed value of $P(r\rightarrow\infty)=c_{p}$ (which as stated earlier, we set at 0.01), we can make the sound speed at infinite radius smaller. We found that for all simulations with sufficiently large sound speeds, the scalar field eventually settles down to a stationary solution. The code was not designed to simulate fields with sound speeds close to zero. Simulated profiles for both $P$ and $S$ at various times are shown in figures \ref{UDMPFig} and \ref{UDMSFig1}. Note that although our initial conditions have a time-like gradient for the scalar field, during the course of the dynamical accretion the gradient became space-like at some regions. This can be seen from figures \ref{PosAPFig}, \ref{PosASFig1}, \ref{UDMPFig}, and \ref{UDMSFig1}, where for the models given in eqn. \eqref{Vik} (for positive $\alpha$) and eqn. \eqref{ghost}, $S$ is larger than $P$ in some regions of spacetime.

\begin{figure}
\includegraphics{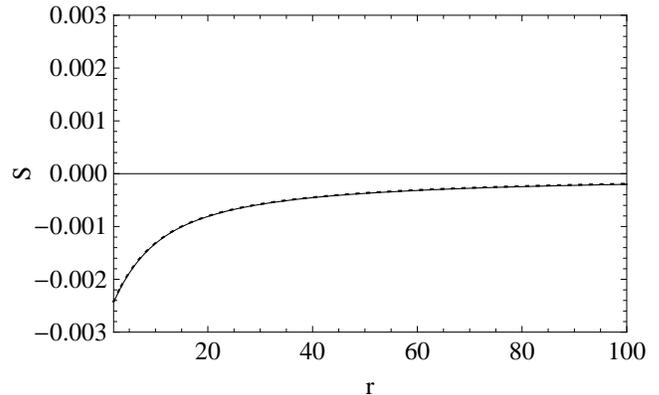}
\caption{The solid line is the radial derivative of the stationary solution for $\phi$ eventually arrived at via simulation of the accretion of the field described by \eqref{ghost}. The dotted line is the radial derivative of $\phi$ for the analytic stationary solution. For this particular plot, we used $c_{p}=0.01$, $a=0.0638943$ (which corresponds to a sound speed at infinite radius having the value $c_{\infty}=\frac{1}{2}$), and $M=1$. The sound horizon has settled down at $r=3.21$. The radial derivative was chosen since the radial derivative of a stationary solution is time independent.}
\label{fig3}
\end{figure}

We then compared the stationary solution we found from the simulation to the analytically derived stationary solution \cite{frolov}. For this model, deriving the analytic solution involves solving a cubic equation (for details see the appendix). A comparison of the analytic solution (evaluated using Mathematica) and the simulated solution at late times is given in figure \ref{fig3}.

\section{Conclusion}

Upon simulating the accretion of two very different types of non-canonical scalar fields onto a black hole, we find that both eventually settle down to a stationary solution. This is reassuring as it suggests that the numerous studies done on the stationary solutions of scalar field models with non-standard kinetic terms correspond to physically viable scenarios as opposed to special situations which cannot be reached dynamically from generic initial conditions.

We find that for a k-essence scalar field with a DBI-type action, if the parameters are chosen such that the field can propagate superluminally, one can send signals from within a black hole as the field approaches this stationary solution. This was confirmed by numerically integrating outgoing null geodesics with respect to the emergent k-essence metric during simulations of this accretion process.

Simulations of a typical ghost condensate action revealed that the stationary solution reached is that derived analytically by Frolov in \cite{frolov}. He showed that this particular solution has an extremely high accretion rate, which puts very strong constraints on the ghost condensate model. However, for the case where the sound speed is zero, the solution is ambiguous (see appendix for further details). Mukohyama has argued in \cite{mukho} that the sound speed should be extremely small on physical grounds, and for corresponding solutions in this limit the accretion rate is reasonable. As our numerical method is not meant to deal with extremely small sound speeds, it would be interesting to see this particular issue resolved with the use of another method. 

\begin{acknowledgements}
We would like to thank Shinji Mukohyama for useful discussions. The work of DG was supported by NSF grant PHY-0855532 to Oakland University. RA and RS are supported by a grant from the U.S. Department of Energy.
\end{acknowledgements}

\newpage

\section*{Appendix A}

Here we present some details concerning the derivation of the stationary solution used in figures \ref{fig1}, \ref{fig2}, and \ref{fig3} that were found by Babichev et. al. \cite{bmv} and Frolov \cite{frolov}. We generalize their solutions to our formalism and present further details relevant to the results presented earlier in this paper. In this appendix we use standard Schwarzschild coordinates ($ds^{2}=-f(r)dt^{2}+f^{-1}(r)dr^{2}+r^{2}d\Omega^{2}$). These coordinates are related those that we use in the main text by equations \eqref{coords1} and \eqref{coords2}.

For a Lagrangian $\mathcal{L}(X)$ that is purely a function of the kinetic variable X, the equation of motion can be written as:
\begin{align}
\nabla_{\mu}(\mathcal{L}_{X}\nabla^{\mu}\phi)=0 \ ,
\end{align}

We use the stationary ansatz\footnote{In \cite{agv} we have shown that this form of the solution directly follows from the requirement of stationarity, c.f. with the so-called delayed field approximation from \cite{Frolov:2002va}}:

\begin{align}
\phi=c_{p}\left(t+r*+\int F(r)dr\right) \ ,
\end{align}
where $r*=\int f^{-1}(r)dr$.
Note that under this ansatz, in the Schwarzschild metric we have:
\begin{align}
\nabla_{\mu}\phi=c_{p}(\nabla_{\mu}t+f^{-1}W\nabla_{\mu}r) \ ,
\end{align}
where we have defined: $W\equiv1+fF$.
\\
We also see that:
\begin{align}
\nabla^{\mu}\phi=c_{p}(-f^{-1}\delta^{\mu}_{t}+W\delta^{\mu}_{r}) \ .
\end{align}
So with this ansatz (applying the general formula for divergence) we find that the equation of motion becomes:
\begin{align}
\frac{1}{r^{2}}\p_{r}(r^{2}\mathcal{L}_{X}c_{p}W)=0 \ .
\end{align}
This gives us the following equation for W:
\begin{align}
c_{1}=r^{2}\mathcal{L}_{X}W \ ,
\end{align}
where $c_{1}$ is a constant. Note from $\nabla_{\mu}\phi$ and $\nabla^{\mu}\phi$ we have that:
\begin{align}
X=\frac{1}{2}c_{p}^{2}f^{-1}(1-W^{2}) \ .
\end{align}
For the Lagrangian given in \eqref{Vik}, we have that 
\\$\mathcal{L}_{X}=(1+\frac{2X}{\alpha})^{-1/2}$, so our equation for W becomes:
\begin{align}
W^{2}=\frac{f+c_{\infty}^{2}-1}{\frac{r^{4}f}{c_{1}^{2}}+c_{\infty}^{2}-1} \ .
\end{align}
In order to ensure that the solution is non-singular on and outside the sound horizon, we choose $c_{1}=\frac{4M^{2}}{c_{\infty}^{4}}$ \cite{bmv}. This uniquely defines $W$ and thus $F$, giving us the closed form for our stationary solution which is plotted in 
figures (\ref{fig1}) and (\ref{fig2}).

For the ghost condensate Lagrangian given in \eqref{ghost}, $\mathcal{L}_{X}=\frac{1}{4}(\frac{X}{a^{4}}-1)$, so the equation for W is:
\begin{align}
r^{2}(\frac{c_{p}^{2}}{2a^{4}}(1-W^{2})-f)W-4c_{1}f=0 \ .
\label{cubic}
\end{align}
Before we go further, consider the horizon at $r=2M$. Because $f[2M]=0$, we know that the equation for $W[2M]$ becomes:
\begin{align}
(1-W[2M]^{2})W[2M]=0 \ .
\end{align}
Thus both $W[2M]=\pm1$ and $W[2M]=0$ are possible solutions. Note that since $W=1+fF$, if $W[2M]=1$, then $F[2M]$ is nonsingular. If $W[2M]=0$ or $W[2M]=-1$, $F[2M]$ is singular and nonphysical.

Upon solving the full cubic equation using Mathematica (as can be expected, these cubic roots are neither brief nor illuminating and are omitted), we chose the solution for $W$ that has $W[2M]=1$. Normally, as was in the previous case, we would choose the value of $c_{1}$ that makes $W[r\rightarrow\infty]=0$, as this prevents $\phi \sim r$ at large radii. However, we find that in the limit $r\rightarrow\infty$, $W$ is independent of $c_{1}$. Furthermore, the cubic root that corresponds to $W[2M]=1$ goes to a non-zero constant as  $r\rightarrow\infty$. However, there is another cubic root which has $W[r\rightarrow\infty]=0$. We can stitch these two separate roots together provided there exists some radius at which the roots become degenerate and coincide with each other.

This occurs for critical values of $W$ and $r$ where the full differential of the cubic equation \eqref{cubic} vanishes. These are the values for $W$ and $r$ where both coefficients vanish in the equation:
\begin{align}
\frac{dC(W,r)}{dW}dW+\frac{dC(W,r)}{dr}dr=0 \ ,
\end{align}
where $C(W,r)=0$ gives the cubic equation defined in \eqref{cubic}.
The constant $c_{1}$ can then be chosen so that these critical values of $W$ and $r$ are solutions to the cubic equation. We can then stitch together the two solutions (one with proper limiting behavior at the horizon and the other with proper limiting behavior at infinite radius) at this critical radius. The resulting solution is well behaved, and is plotted in figure (\ref{fig3}).

One exceptional case where this procedure is ambiguous is when we have $c_{\infty}=0$ or equivalently, $c_{p}^{2}=2a^{4}$. In this case as $r\rightarrow\infty$, all three cubic roots have $W[2M]=0$. Thus, there is no longer an obvious condition on $c_{1}$.

\bibliographystyle{utphys}

\end{document}